# Ab-initio investigation of transition metal dichalcogenides for the hydrogenation of carbon dioxide to methanol

Avaneesh Balasubramanian [a], Pawan Kumar Jha [b], Kaustubh Kaluskar [c], Sharan Shetty [c], and Gopalakrishnan Sai Gautam *[b]

[a] Indian Institute of Science Education and Research, Pune 411008, India

[b] Department of Materials Engineering, Indian Institute of Science, Bengaluru 560012, India

[c] Shell India Markets Pvt. Ltd., Bengaluru 562149, India

*E-mail: saigautamg@iisc.ac.in

## Abstract

We computationally investigate the catalytic potential of $MoSe_2$, $WS_2$, and $WSe_2$ nanoribbons and nanosheets for the partial hydrogenation of $CO_2$ to methanol by comparing their electronic, adsorption, and defect properties to $MoS_2$, a known thermo-catalyst. We identify Se-deficient $MoSe_2$ (followed by $WSe_2$) nanosheets to be favorable for selective methanol formation.

# Introduction

Transforming $CO_2$ into valuable chemicals is an effective strategy to reduce anthropogenic $CO_2$ and promote a circular carbon economy. One way to transform $CO_2$ is to catalytically convert it into methanol (MeOH), which serves as a precursor for renewable fuels like sustainable aviation fuel and other useful compounds such as formaldehyde [1-6]. However, several catalytic processes to produce MeOH from $CO_2$ suffer from low conversion and stability or take place at high temperatures making it economically unattractive.

The thermo-catalytic route to convert $CO_2$ to MeOH has been well studied over the years, with Cu/ZnO being one of the most studied catalytic systems [7]. While the mechanism for the $CO_2$ conversion on Cu/ZnO is not fully understood, oxygen vacancies on the surface seem to play a significant role [8]. Recently, a study has reported that the in-plane sulphur vacancies of $MoS_2$ nanosheets are highly selective (94.3% selectivity at 12.5% conversion) towards partial reduction of $CO_2$ to MeOH at low temperatures, with support from spectroscopic and structural characterization and theoretical calculations [9]. Other transition metal dichalcogenides (TMDC) like $MoSe_2$, $WS_2$, and $WSe_2$ closely resemble $MoS_2$ in their crystal structures ($P6_3/mmc$ space group), bulk electronic structures, and overall chemistries [10], suggesting the potential of TMDCs besides $MoS_2$ as potential thermo-catalysts for $CO_2$ conversion.

For instance, computational and experimental studies on $MoSe_2$, $WS_2$, and $WSe_2$ demonstrate the similarities in the surface activity of these compounds, attributed to the similarities in their electronic structures [11-14]. Additionally, these materials are well-known 2D catalysts, with applications in the hydrogen evolution reaction (HER) [15]. Notably, a recent study [16] has experimentally reported a 93% selectivity at a 9.7% conversion of $CO_2$ by C-doped $MoSe_2$ by the thermo-catalytic approach, indicating promising of such TMDCs for $CO_2$ conversion. However, detailed insights on the reaction mechanism including computations, which would allow further optimization of such catalysts, was not part of the study [16]. Moreover, to the best of our knowledge, there have been no systematic computational studies comparing and validating the activity of chalcogenide-vacant TMDCs (of composition $MX_2$ with M = Mo, W and X = S, Se) for $CO_2$ conversion, motivating this work [6, 17, 18].

In this work, we use density functional theory (DFT) [19, 20] calculations to compare, with $MoS_2$, the electronic structures, and defective and adsorption energetics of single, double, and triple chalcogenide-vacant TMDC nanosheets and nanoribbons, namely $MoSe_2$, $WS_2$ and $WSe_2$, to explore their potential as catalysts for the partial hydrogenation of $CO_2$ to MeOH. Since morphology often plays a crucial role in the efficiency of TMDC catalysts [21], we consider all TMDCs in this work to be of nanosheet and nanoribbon morphologies. Specifically, we compute the electronic density of states (DOS), vacancy formation energies, and $CO_2$ and MeOH adsorption energies for all four TMDCs considered. We compare the calculated properties of $MoSe_2$, $WS_2$, and $WSe_2$ with $MoS_2$ to examine the similarities and arrive at combinations of candidate material, vacancy content, and morphology that can exhibit partial reduction of $CO_2$ to a similar extent as $MoS_2$. While we find all TMDCs considered to exhibit similar qualitative trends in calculated properties compared to $MoS_2$, we observe in-plane double-Se-vacant $MoSe_2$ nanosheets (and to an extent, $WSe_2$ nanosheets) to be the most promising for $CO_2$ reduction due to its close similarities to the $MoS_2$ electronic structure, defect, and adsorption energetics. We hope that our work further instigates work in the development of TMDC catalysts for $CO_2$ reduction.

## Structures

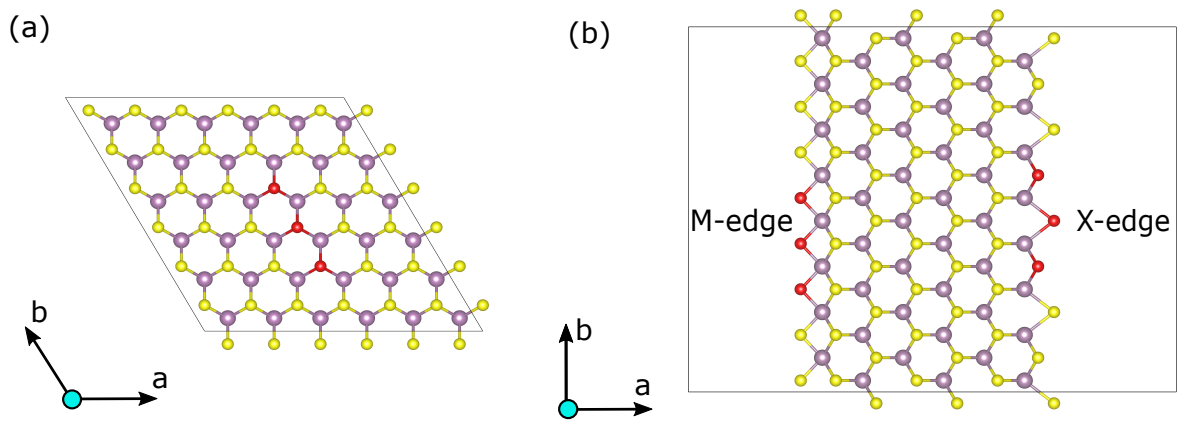

**Figure 1**: The structures of TMDC nanosheets (panel a) and nanoribbons (panel b), with the yellow and purple spheres denoting X (S,Se) and M (Mo, W) atoms, respectively. Red spheres in both panels denote the X-atoms that are removed sequentially according to their numerical labels to form the single, double and triple vacancies. We consider X-vacancies to form on the in-plane sites of the nanosheets, and at the M- and X-edges of the nanoribbons.

Figure 1 displays the structures used for modelling the nanosheets (panel a) and nanoribbons (panel b) of the TMDCs considered in this work. Nanosheets are single layers with M atoms (purple spheres) arranged in a plane, bonded to X atoms (yellow spheres) 'above' and 'below' the plane, forming dimers. Following [9], we generate the in-plane vacancies of adjacent X atoms 'above' the M-plane of the nanosheets, in the sequence shown by the numerical labels in Figure 1, resulting in single, double, and triple X-vacant nanosheets. In the nanoribbons, the M-edge is formed by capping the M-atoms at the edge with single X atoms in the M-plane, while the X-edge is formed by alternating X dimers and single planar X atoms bonded to M. We generate the X-vacancies on both the M- and X-edges of the nanoribbons sequentially, in accordance with the numerical labels. More details on the structures used is provided in the electronic supporting information (ESI).

# Results

## Band gaps and defect formation energies

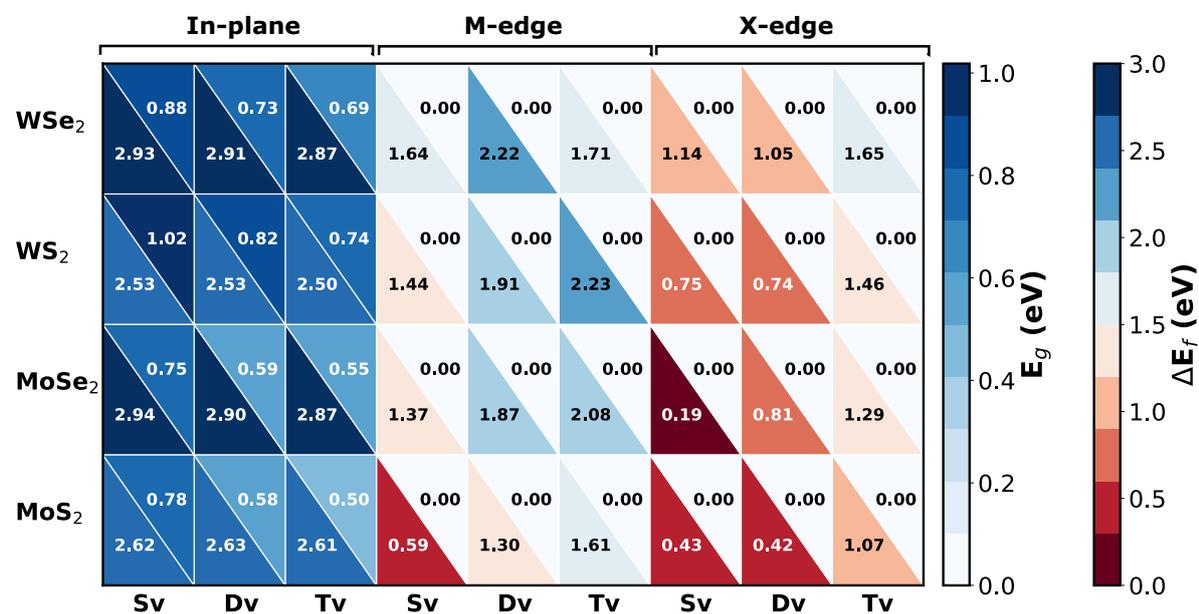

**Figure 2**: Heatmap of calculated band gaps ($E_g$) in the upper triangles of each cell and X-vacancy formation energy ($\Delta E_f$) in the lower triangles, for each vacancy at a TMDC. Sv, Dv, and Tv indicate single, double, and triple X-vacant configurations. In-plane refers to nanosheet morphology, while M- and X-edges are part of nanoribbon. $\Delta E_f$ are reported in eV per X-vacancy formed.

The catalytic activity of a material is primarily influenced by its electronic structure and the resultant alignment of bands to the electronic states of the reactant [22]. Figure 2 plots the calculated band gap ($E_g$ in eV), based on DOS calculations (see Figures S1-S6 of the ESI) within the upper triangles of each cell for single (Sv), double (Dv), and triple (Tv) X-vacant TMDCs. Metallic states are denoted by white triangles ($E_g = 0$ eV), with dark blue triangles indicating $E_g$ approaching 1 eV. 'In-plane' denotes calculations done with the nanosheet morphology while 'M-edge' and 'X-edge' signify the nanoribbon.

DOS calculations for X-vacant MoSe$_2$, WS$_2$, and WSe$_2$ nanosheets reveal that it lacks electronic states at the Fermi level (i.e., non-metallic), like MoS$_2$. Importantly, the $E_g$ of the nanosheets of all TMDCs decrease (by ~0.2-0.3 eV) as more in-plane X-vacancies are introduced. X-vacant TMDC nanoribbons, with the X-vacancies at both the M- and X-edges are metallic, similar to our observation in $MoS_2$ as well (Figure 2 and S5). As observed in [9], the presence of electronic states at the Fermi level in the nanoribbons might hinder the desorption of MeOH formed during hydrogenation from the X-vacant sites, reducing the selectivity of the TMDC nanoribbons towards MeOH release.

Compared to $MoSe_2$ and $WSe_2$, the X-vacant $WS_2$ nanosheets have fewer states/eV at the band edges (Figure S3), suggesting slower kinetics towards $CO_2$ reduction. Importantly, $MoSe_2$ exhibits the most similar electronic structure to $MoS_2$, suggested by the similarities in the calculated DOS for the defective nanosheet configurations (Figure S3) and the resultant $E_g$ only differing by ~0.05 eV (Figure 2). This similarity implies that the alignment of the electronic states of in-plane X-vacant $MoS_2$, which are the most active sites towards MeOH formation from $CO_2$ [9, 23], towards $CO_2$ and the reaction intermediates could also be observed for the Se-vacant $MoSe_2$ nanosheets, making it a possible catalytic candidate. Indeed, the DOS of $CO_2$ adsorbed at the X-vacant TMDC nanosheets, particularly $MoS_2$ and $MoSe_2$ (Figure S6), suggests the alignment of the O $p$ states of $CO_2$ near the M $d$-states, which can facilitate the donation of electrons from the $d$-orbital (HOMO) of the M atoms that are adjacent to X-vacancies to the LUMO of $CO_2$, facilitating hydrogenation across the C-O bonds.

In addition to electronic structure, the presence of a sufficient number of active sites (the number of X-vacancies in TMDC surfaces and edges in this work) crucially determines a catalyst's effectiveness. To assess the feasibility of generating X-vacancies in pristine nanoribbons and nanosheets, we compute the vacancy formation energies ($\Delta E_f$ in eV, lower triangles in Figure 2) for Sv, Dv, and Tv TMDC nanosheets (in-plane) and nanoribbons (M-edge and X-edge). A low (closer to zero or brown triangles in Figure 2) vacancy formation energy indicates a higher equilibrium concentration of X-vacancies at the TMDC surface, which would result in an increased number of active sites for $CO_2$ and $H_2$ binding, thereby enhancing the reaction process and positively impacting the catalytic performance.

As seen in Figure 2, the in-plane S-vacancy $\Delta E_f$ on $MoS_2$ nanosheets are similar to what is reported in [9], with the exception that we find it easier to form S-vacancies at the S-edge than at the Mo-edge in the nanoribbon, which can be attributed to the higher availability of S in the S-edge resulting in easier S-vacancy formation. Like $MoS_2$, the other TMDCs show lower $\Delta E_f$ at the X-edge than at the M-edge as well. The X-vacancy $\Delta E_f$ at the in-plane sites of the $MoSe_2$ and $WSe_2$ nanosheets are higher than those of $MoS_2$ (by ~0.25-0.32 eV), with $WS_2$ showing lower $\Delta E_f$ (by 0.1 eV). In case of the nanoribbons, all TMDCs exhibit larger $\Delta E_f$ than $MoS_2$. The higher $\Delta E_f$, especially in $MoSe_2$ and $WSe_2$, indicates that the formation of vacancies in these nanoribbons and nanosheets will be less favourable. Given that the in-plane $\Delta E_f$ differ by ~0.3 eV in $MoSe_2$ (and $WSe_2$) from the corresponding $MoS_2$ configurations, we can expect a difference in equilibrium concentration of the in-plane vacancies in nanosheets (active sites) to be ~three orders of magnitude at an operating temperature of ~473 K.

**$CO_2$ adsorption energies:**

The affinity of the active site(s) towards the reactant ($CO_2$ here) plays an important role in catalysts' performance. The reactant must bind well to the active site to undergo the reaction and facilitate electron transfer. However, excessive binding could lead to the poisoning of the catalyst's active site. Figure 3 quantifies the $CO_2$ adsorption energies at different vacancy concentrations (indicated by the colours of each symbol) for the TMDCs considered (signified by shapes of each symbol) in both the nanosheet (in-plane) and nanoribbon (M- and X-edge) morphologies. Since $H_2$, which is the co-reactant of $CO_2$ for reduction to MeOH, is a significantly smaller molecule compared to $CO_2$ and is typically weakly adsorbed (or physisorbed) at the active sites, we neglect its presence for the calculation of $CO_2$ adsorption energy. Note that a site with $CO_2$ adsorption energy that is similar to the active site (in-plane X-vacancies) of $MoS_2$ is expected to show optimal affinity towards $CO_2$ for partially hydrogenating it. The $CO_2$ molecule disintegrates upon adsorption in the Dv M-edge sites of $WS_2$ and $WSe_2$, which is why they are not shown in Figure 3.

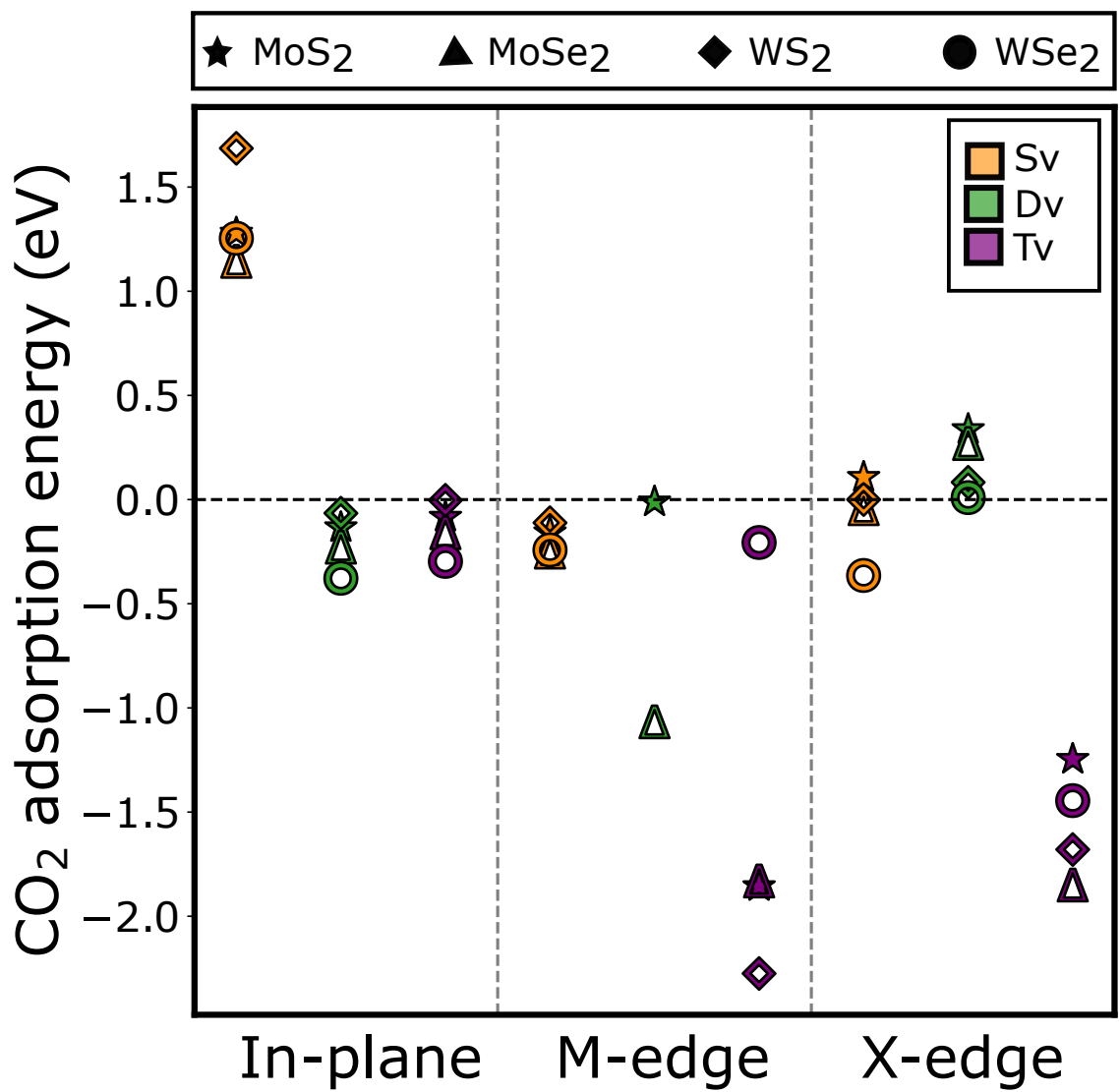

**Figure 3**: $CO_2$ adsorption energies in TMDC nanosheets (in-plane) and nanoribbons (M- and X-edge) with X-vacancies. Configurations containing single (Sv), double (Dv), and triple (Tv) vacancies are indicated by orange, green, and purple markers, respectively. Datapoints corresponding to $MoS_2$, $MoSe_2$, $WS_2$, and $WSe_2$ are indicated by hollow stars, triangles, diamonds, and circles, respectively.

As for the S-vacant MoS$_2$ nanoribbon, we find that all the Mo-edge sites and triple-S-vacant S-edge sites show negative CO$_2$ adsorption energies (Figure 3), indicating strong binding. Notably, the X-edge Tv structures show the lowest CO$_2$ adsorption energies in the corresponding TMDC nanoribbons, indicating excessive binding and potential poisoning of the X-vacant sites. In contrast, CO$_2$ adsorption is energetically unfavourable on all TMDC nanosheets on the Sv sites, indicating that clustered X-vacancies (i.e., Dv and Tv configurations) are required to kickstart the CO$_2$ reduction. Similar to MoS$_2$, we find the adsorption energies of CO$_2$ to be the lowest (most negative) on Dv sites, among the nanosheets. Therefore, like in MoS$_2$ [9], an appreciable CO$_2$ adsorption can also be expected from the other double-X-vacant TMDC nanosheets. Comparing the adsorption energies of the in-plane Dv sites, we find MoSe$_2$ and WSe$_2$ sheets to show better CO$_2$ adsorption than MoS$_2$ and WS$_2$.

**MeOH adsorption energies:**

To verify the ability of TMDC nanosheets and nanoribbons to release MeOH once it is formed and not stay adsorbed and further hydrogenate to methane, we calculate the adsorption energy of MeOH on the X-vacant nanoribbons and nanosheets and plot them in Figure 4. A higher (more positive) adsorption energy implies quick release and therefore higher selectivity to MeOH. From Figure 4, we observe that the adsorption energy of MeOH is higher at the in-plane vacant sheets compared to the M-edge (and X-edge) vacant ribbons for all the TMDC candidates, signifying a higher selectivity towards forming MeOH at the in-plane sites, which is also observed for MoS$_2$ in [9]. In all the TMDC ribbons, the Dv site shows the lowest MeOH adsorption energy at the M-edge, and the Tv site shows the lowest MeOH adsorption energy at the X-edge. Compared to the in-plane Dv site of MoS$_2$, the corresponding sites of the other TMDCs show more positive MeOH adsorption energies, with MoSe$_2$ showing the most positive value. Hence, among the TMDCs considered, the active site of in-plane Dv of MoSe$_2$ desorbs MeOH better than MoS$_2$ and hence should be more selective to MeOH formation and less prone to complete hydrogenation of CO$_2$ to methane.

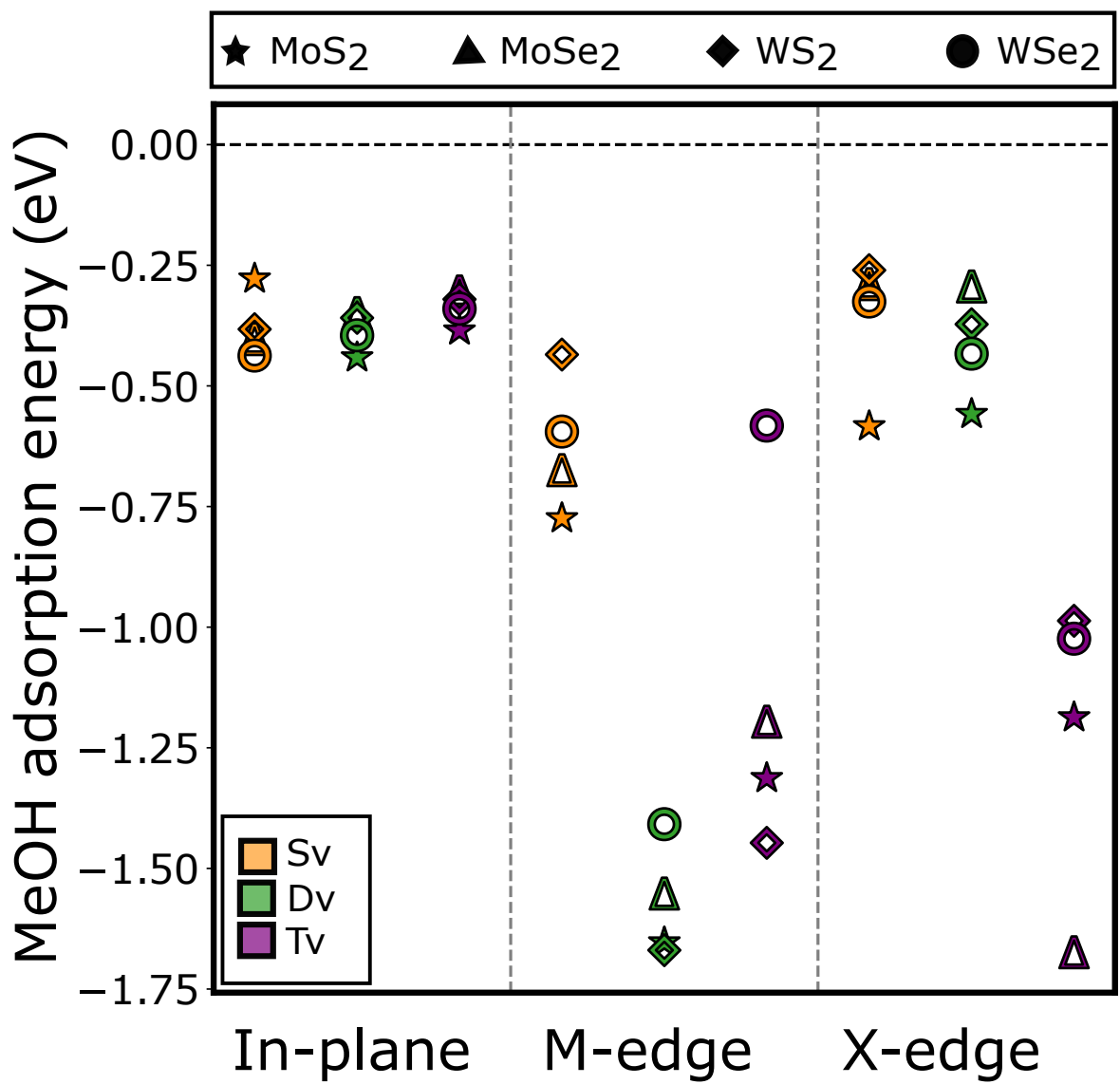

**Figure 4**: MeOH adsorption energies in X-vacant TMDC nanosheets and nanoribbons considered. Notations in the figure are identical to those used in Figure 3.

## Discussion

Our results indicate that the TMDC ($MoSe_2$, $WS_2$ and $WSe_2$) nanoribbons and nanosheets considered show similar electronic (as seen in the electronic DOS), defect properties (as seen in vacancy $\Delta E_f$) and adsorption thermodynamics (adsorption energies of $CO_2$ and MeOH) to $MoS_2$. However, we find that Se-vacant $MoSe_2$ nanosheets are the most similar to the S-vacant $MoS_2$ nanosheets. The similarity between $MoSe_2$ and $MoS_2$ can be observed from the similar DOS and $E_g$ of the X-vacant $MoSe_2$ and $MoS_2$ structures, better adsorption of $CO_2$ at the in-plane Dv sites of $MoSe_2$ and more positive adsorption energies of MeOH at $MoSe_2$ than at $MoS_2$. Since the studies in [9] and [23] suggest that the in-plane vacancies of $MoS_2$ are the most active sites for partial hydrogenation of $CO_2$ to MeOH, and we find that the in-plane Se-vacant $MoSe_2$ show similar properties to those of $MoS_2$, we predict $MoSe_2$ nanosheets to be the most promising candidate among the TMDCs considered for further experimental and computational studies on $CO_2$ reduction. Additionally, $WSe_2$ nanosheets are comparable to $MoSe_2$ in terms of their adsorption energy of $CO_2$ and relatively higher adsorption energy with MeOH, albeit with larger $E_g$. Thus, $WSe_2$ nanosheets can be considered for $CO_2$ reduction to MeOH as well.

We predict that the M-edge X-vacant nanoribbons would be less selective for MeOH during hydrogenation due to the presence of states at the Fermi level in their electronic structure (Figure S5, which is also reflected in their significant binding of MeOH (Figure 4). While we predict the X-edge X-vacant nanoribbons to be metallic, the presence of states at the Fermi level of these nanoribbons does not necessarily imply poor selectivity towards MeOH, especially in the Sv and Dv configurations. Hence, studies utilising by microkinetic and other experimental adsorption observations are required to ascertain the selectivity of the X-edge of the nanoribbons. However, given the overall electronic structure, defective and adsorption energetics, we do not expect any of the TMDC nanoribbon morphologies considered in this work to be of significant utility for $CO_2$ reduction to MeOH.

## Conclusion

Utilising anthropogenic $CO_2$ by transforming it into useful chemicals, fuels, and precursors, such as MeOH, is an important way of ensuring a circular carbon economy. In this study, we used DFT calculations to quantify the electronic structure, vacancy $\Delta E_f$, and $CO_2$ and MeOH adsorption energies of four TMDCs to explore their utility as thermo-catalysts for $CO_2$ reduction to MeOH. We considered different concentrations of X-vacancies (Sv, Dv, and Tv) and two different morphologies (nanosheets and nanoribbons) of $MoS_2$, $MoSe_2$, $WS_2$, and $WSe_2$. Motivated by recent studies [9, 23] that have reported $MoS_2$ to show a high degree of selectivity for $CO_2$ conversion to MeOH, we analysed similarities in the calculated properties among the TMDCs to identify potential candidates. Importantly, we found $MoSe_2$ nanosheets (with Dv of Se) to exhibit $E_g$, vacancy $\Delta E_f$, and adsorption energies that resemble closest to $MoS_2$, identifying $MoSe_2$ to be the most promising among the TMDCs considered, followed by $WSe_2$. We hope that our work enables the identification of other 2D materials as possible catalysts for selective $CO_2$ reduction.

**Acknowledgment**

The authors acknowledge financial support from Shell India Markets Private Limited. The authors gratefully acknowledge the computational resources of the super computer 'PARAM Pravega' provided by Super Computer Education and Research Centre (SERC), IISc. Kaustubh

Kaluskar and Sharan Shetty would like to thank Joost Smits and Sander van Bavel (Shell Global Solutions International B. V.) for fruitful discussions.

**Conflicts of interest**

There are no conflicts to declare.

**Data availability**

The computational data supporting this study is openly available at our [GitHub](GitHub) repository.